# Science Gateway for Distributed Multiscale Course Management in e-Science and e-Learning — Use Case for Study and Investigation of Functionalized Nanomaterials

Yu. Gordienko*,**, S. Stirenko**, O. Gatsenko* and L. Bekenov*

* G.V.Kurdyumov Institute for Metal Physics, National Academy of Sciences, Kyiv, Ukraine
** National Technical University of Ukraine "KPI"/ High Performance Computing Center, Kyiv, Ukraine
e-mail: gord@imp.kiev.ua

*Abstract* - The current tendency of human learning and teaching is targeted to development and integration of digital technologies (like cloud solutions, mobile technology, learning analytics, big data, augmented reality, natural interaction technologies, etc.). Our Science Gateway (http://scigate.imp.kiev.ua) in collaboration with High Performance Computing Center (http://hpcc.kpi.ua) is aimed on the close cooperation among the main actors in learning and researching world (teachers, students, scientists, supporting personnel, volunteers, etc.) with industry and academia to propose the new frameworks and interoperability requirements for the building blocks of a digital ecosystem for learning (including informal learning) that develops and integrates the current and new tools and systems. It is the portal for management of distributed courses (workflows), tools, resources, and users, which is constructed on the basis of the Liferay framework and gUSE/WS-PGRADE technology. It is based on development of multi-level approach (as to methods/algorithms) for effective study and research through flexible selection and combination of unified modules ("gaming" with modules as with LEGO-bricks). It allows us to provide the flexible and adjustable framework with direct involvement in real-world and scientific use cases motivated by the educational aims of students and real scientific aims in labs.

## I. INTRODUCTION

The current tendency of human learning and teaching is targeted to development and integration of digital technologies like cloud solutions, mobile technology, learning analytics, big data, augmented reality, natural interaction technologies, etc. But the available e-Science and e-Learning frameworks (for example, in Massive Open Online Courses (MOOCs) [1] and OpenCourseWare [2] initiatives, like TIMMS [3], SlideWiki [4], Wikiversity [5], edX [6], Coursera [7], et al.) are: (a) very heterogeneous (the available learning content is quite different in various institutions), (b) often far away from the real-life learning situations, (c) in general, static like stationary web-pages with online access from desktop PCs only, (d) not flexible to the very volatile demands of end users (students, pupils, ordinary people) and real-world, (e) not-adjustable by complexity, duration, and range (e.g. for taking some modules, but not the whole course), especially in the life-long learning or vocational training, (f) the feedback is not available about the quality, value, and necessity of some parts of the course or combinations of some modules.

The paper is aimed on the description of the approach to close cooperation among the main actors in learning world (teachers and students) with industry and academia. Here, the new frameworks and interoperability requirements are considered for the building blocks of a digital ecosystem for learning (including informal learning) that develops and integrates the current and new tools and systems.

## II. BACKGROUND

A vast amount of e-Science and e-Learning frameworks briefly enlisted in "Introduction" is published online to make their educational content accessible to a wide range of audiences. Many profound researches were dedicated to their estimation in the view of awareness of these courses among users and increase of popularity of these frameworks [7-9]. However, from a subjective experience of many users, online courses are mainly dull, outdated, and/or non-attractable. As it was observed in one of the researches [8]: ~28% of the courses are indeed open, ~12% are available in a language other than English, ~16% are available in a format facilitating reuse and repurposeability, ~30% were updated in the last 3 years, and <50% comprise self-assessment questions, only ~65% have at least one example and one illustration, ~60% have been objectively determined as low attractive, and only 10% have been categorized as highly attractive. In the context of this investigation (on 100 various online courses), the immediate conclusion is that the quality of these e-Learning courses and frameworks should be improved significantly, especially in the sense of better attractiveness, feedback (more interactive examples) and applicability (more practical use cases).

The work presented here was partially funded by EU FP7 SCI-BUS (SCIentific gateway Based User Support) project, No. RI-283481, EU TEMPUS LeAGUe (A Network for Developing Lifelong Learning in Armenia, Georgia and Ukraine) project No. 543839-TEMPUS-1-2013-1-SE-TEMPUS-SMHES, and partially supported in the framework of the State Targeted Scientific and Technical Programs in 2013-2015.

A possible solution for improving the quality can be found by leveraging the collaboration and sharing experience and e-Learning courses/frameworks. Several platforms (like SlideWiki [4] or Wikiversity [5]) were mentioned in "Introduction", which can be used to support the collaborative creation and usage of e-Learning courses by a wide range of authors (teachers, developers, providers) and users (students, ordinary people) communities.

In addition to this, the big potential of the recent "science gateway" (SG) ideology can be used [10]. In general, SG means a user-friendly interface between users (for example, scientists or scientific communities), various software tools and hardware resources, usually distributed computing infrastructures (DCIs). The aim SGs is to decrease the time and effort spent in setting up the environment and DCI for a specific scientific application. By means of SGs researchers can concentrate their efforts on their scientific goals and thus spending less time and effort on assembling and managing the required DCI components. The most important aspects of SG are: a simplified intuitive graphical user interface (GUI) that is highly tailored to the needs of the given scientific community; a smooth access to national and international computing and storage resources; collaborative tools for sharing scientific data. Nowadays various SG frameworks exist (for example, ASKALON [11], KNIME [12], MOTEUR [13], gUSE/WS-PGRADE [14]), which use different enabling components and technologies: web application containers (Tomcat, Glassfish, etc.), portal or web application frameworks (Liferay, Spring, Drupal, etc.), database management systems (MySQL, etc.), and workflow management systems. Here, our SG for distributed multiscale course management in e-Science and e-Learning, which is based on gUSE/WS-PGRADE framework, is described. The detailed description of SG ideology on the basis of gUSE/WS-PGRADE (https://guse.sztaki.hu) can be found elsewhere [14]. In short, WS-PGRADE portal is a web based front end of the gUSE infrastructure. gUSE (grid and cloud User Support Environment) is a framework that provides the user-friendly access to DCI resources. SG contains the workflow-oriented graphical user interface (GUI) to create and manage workflows for tasks executed on various DCI resources. The DCI resources can include PCs, clusters, service grids, desktop grids, clouds [15-17].

### III. SCIENCE GATEWAY FOR COURSE MANAGEMENT

The aim of the approach is to increase attractiveness of e-Learning courses by direct involvement in design of learning process, feedback by more interactive examples and applicability by more practical use cases. It could be performed by development of multi-level approach (methods/algorithms) for effective learning through flexible selection and combination of unified modules ("gaming" with modules as with LEGO-bricks) under SG. This can provide the flexible and adjustable framework with the elements of direct involvement in real-world and scientific use cases. It should be motivated by the educational aims of students and real scientific aims in labs [16].

#### A. Objectives

The objectives of the approach are to design, create and implement:
- smart learning environments providing students with adaptive and personalized learning and assessment through multi-modal/multi-sensory interaction technologies and advanced interfaces;
- interactive learning systems with more efficient and natural ways of delivering content to users especially for the real scientific tasks;
- technologies to match and operate multiple applications in real environments, while research is expected to be based on freely available and re-usable tools and resources.

#### B. General Structure and Main Components

Due to SG technology several sets of modules can be linked together by the end users (teachers or students) along the desirable track, which can form the course customized by the end users, for example, for molecular dynamics (MD) simulations in computational physics (Fig. 1). To support such linking some basic requirements should be formulated and satisfied for wrapping and connecting the modules. They should be considered as meta-information (Table 1), which are obligatory for successful combination of various modules (mainly for their compatibility). In addition to this, the similar vocabulary should be used among designers and creators of the modules, which is the simple task for the developers from the same institution, but could be headache for combination of modules from various institutions.

The main components of the approach are: portal(s), portlets(s), content, tools, users, and their repositories. "Portal" is assumed to be collection of components like tools, users, and courses (workflows). The courses could be installed as the separate portlets. It is the technological backbone constructed on the basis of the Liferay framework with customizable scientific gateways (gUSE/WS-PGRADE) based on workflows. "Content" is a set of available "legacy" passive modules and new

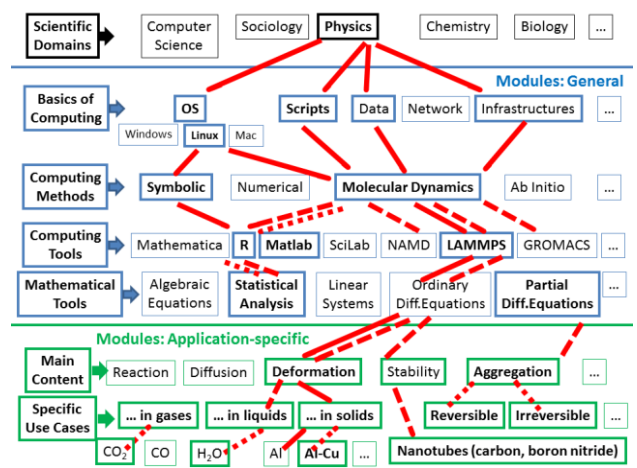

Figure 1. Example of the multiscale course hierarchy in materials science: beginner ("macro") tracks (solid lines), advanced ("mini") tracks (dash lines), expert ("micro" and "nano") tracks (dot lines)

active nano-, micro-, mini-, and macro- modules. The legacy modules should be "wrapped" by standardized meta-information about them to facilitate their combination and construction of workflows (customized tools). The legacy passive modules are assumed to be some sets of web-pages (with multimedia content), which can be enclosed in the portlets and attached to portals. The new active modules are based on the so-called "Virtual Labs" [16], i.e. software workflows (i.e. domain-specific applications connected to some research infrastructures) and enclosed in portlets, which can be attached to portals also. "Repositories" are considered as collections of content in the shape of multiscale modules (like macro-, mini-, micro-, and nano- courses) in the shape of workflows and portlets, which can be uploaded/downloaded and attached to portals.

### C. Meta-information

The compatible meta-information for modules (Table 1) will be necessary for successful combination and compatibility of various modules. The information about "Previous module(s)" is considered as an input socket, i.e. a list of the available modules, which are pre-requisites (obligatory) to use (learn) this module. In similar way "Next module(s)" plays the role of an output socket, i.e. a list of the available modules, which can be used as continuation of this module. "Alternatives" is a list of the available modules, which can be selected as alternative versions of this module. "Categories" describes the place of the current module in the hierarchy of learning disciplines. "Complexity" is a subjective measure of the relative efforts that should be applied by students (it could be used to estimate grades, points, costs, and learning curve). "Scale" is a relative level of the module, i.e. macro-, mini-, micro-, or nano- level. "Duration" is a unit of the module length, which can be measured by time (hours, days, etc.) or lessons. "Workload" is a subjective estimation of time needed for studying the proposed content. "Exercises" are some questions and tasks with answers. "Keywords" are tags, which can be used for search of this module. "Languages" is a measure of universality of the content, i.e. multilingual content is desirable, but English should be obligatory for universal compatibility. Addition of any other languages will increase usability at a local geographical scale. "Rating" is a subjective estimation of the module attractiveness and usefulness, i.e. its quality. "Certificate" is a proof of any formal recognition marks. "Price" is a possible cost for this module (which is usually related with certificate costs, tuition fee, etc.)

### IV. USE CASES FOR STUDY AND INVESTIGATION OF FUNCTIONALIZED NANOMATERIALS

Here we demonstrate the capabilities of this approach applied for MD simulations and further data post-processing on the basis of LAMMPS package for MD simulation [18], and other packages like R for statistical analysis [19], Pizza.py Toolkit for manipulations with atomic coordinates files [20], AtomEye for visualization of atomic configurations [21], debyer for simulation of X-ray diffraction (XRD), and neutron diffraction (ND) analysis (https://code.google.com/p/debyer/), etc.

These use cases (Fig. 2) were targeted on MD simulation of nanocrystals under uniaxial tension (Fig. 3a) and functionalized nanomaterials like carbon nanotubes (Fig. 4a). Their user communities in the field of nanotechnologies include the user groups (teachers and students) from academy (G.V.Kurdyumov Institute for Metal Physics, National Academy of Sciences) and education (National Technical University of Ukraine "KPI"). MD simulation workflow (Fig. 2) was created and used for investigation of relaxation behavior of nanocrystals for different nanocrystals (Al and Al-Cu), physical conditions (tensile rate, size and orientation of the nanocrystals) and methodological parameters

TABLE I. EXAMPLE OF META-INFORMATION

| Category | Format | Example |
|---|---|---|
| Title | String | MD Simulation of Metal Nanocrystals under Deformation |
| Previous module(s) | List of strings | MD Simulation of Metal Nanocrystals |
| Next module(s) | List of strings | MD Simulation of Defect Evolution in Al-Cu Alloys with Nanoinclusions |
| Alternatives | List of strings | MD Simulation of Non-metal Solids |
| Categories | List of strings | Physics:Computational Physics |
| Complexity | Number (1-5) | 4 |
| Scale | String from the set (macro, mini, micro, nano) | Mini |
| Duration | Number (in weeks) | 2 (weeks) |
| Workload | Number (per week) | 8-10 (hours/week) |
| Exercises | Number | 5 |
| Keywords | List of string | MD, metal, alloy, Al-Cu, |
| Languages | List of string | English, Ukrainian |
| Rating | 1-5 stars | 4 |
| Certificate | String (Yes/No) | No |
| Price | Number (in $) | 0 |

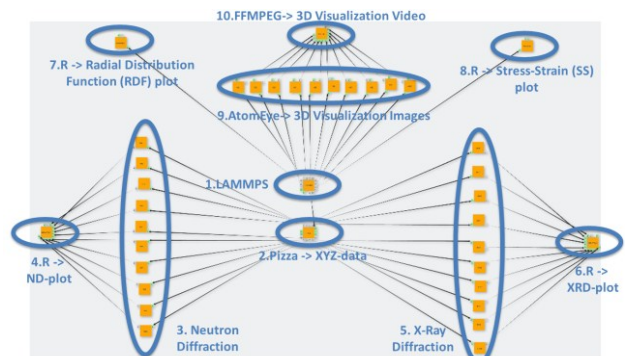

Figure 2. Example of workflow for MD simulations of functionalized nanomaterials. It is created by means of WS-PGRADE package: tools (yellow crates) inside the modules (thick ovals) with connecting links (thin lines)

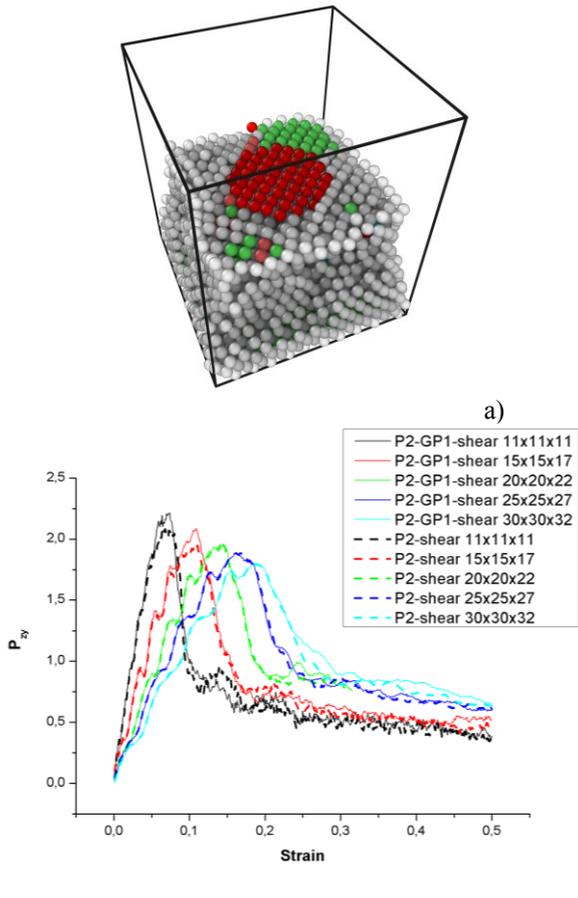

Figure 3. Examples of output data for MD simulations of functionalized nanomaterials: (a) 3D visualization by AtomEye: cross-section of nanocrystal with Cu-nanodisk (Guinier-Preston zone noted by red color), (b) post-processing by R environment: stress-strain plot

(potential type, boundary conditions, and others). The actually used workflow (Fig. 2) includes several subset workflows, which can be considered as separate tracks on the basis of separate modules (crates in SG terminology). The simulations included the following subset workflows: 1) LAMMPS + R-package — for MD simulations and plotting some post-processed values, like stress-strain dependencies; 2) LAMMPS + R-package + AtomEye + FFMPEG — for the same operations + visualization of intermediate states of atoms and final video visualization; 3) LAMMPS + R-package + AtomEye + FFMPEG + debyer — for the same operations + processing data for X-ray and neutron diffraction and their plotting, like XRD-plots. From the physical point of view it was shown that nanocrystal size and complex nanoscale defect substructure influence the strength properties of Al-Cu alloys (Fig. 3b). These results confirm and illustrate the known experimental data described in [22-24].

The potential for variability of this approach can be illustrated by possibility to change input/output scripts (for LAMMPS, AtomEye, and R crates-components) and parameters (for LAMMPS, AtomEye, FFMPEG, and debyer crates). By change of these scripts and parameters the workflow can be applied for the quite different use case, namely, for MD simulations of the nanosubstrates functionalized by carbon nanotubes (Fig. 4a).

As it schematically shown in Fig. 1 the general backbone structure (Fig. 2) can be used in the shape of some subsets for various scientific domains (Physics, Materials Science, Chemistry, Biology, etc.). These subsets can include various sets of modules (user can select the modules connected and emphasized by red color).

Among general modules, for example, "Basics of computing" set could include the crates with some executables dedicated to various computing aspects (basic operations in Linux, Windows, Shell Scripts, Data Operations, Network Basics, ...). "Computing Tools" set can comprise the correspondent binaries (Octave, R, SciLab, Matlab, Mathematica) and input/output scripts for them. "Mathematical Tools" can include static or dynamic content dedicated to some basic and more complex mathematical approaches (like Algebraic Equations, Linear Systems, Ordinary Differential Equations, Partial Differential Equations, etc.). The similar content can be in included "Computing Methods" set (like Symbolic Computation, Numerical Computation, Molecular Dynamics, Ab Initio).

Among application-specific modules, for example, "Main Content" can be related with the specific processes: physical (Diffusion, Aggregation, Reaction, Coalescence, etc.), mathematical (Random Walk, Fractals, etc.), biological, financial, geological, astronomic, and others.

Finally, "Specific Use Cases" can be concentrated on the more practical examples, which are illustrated in this section (Fig. 2 and 3).

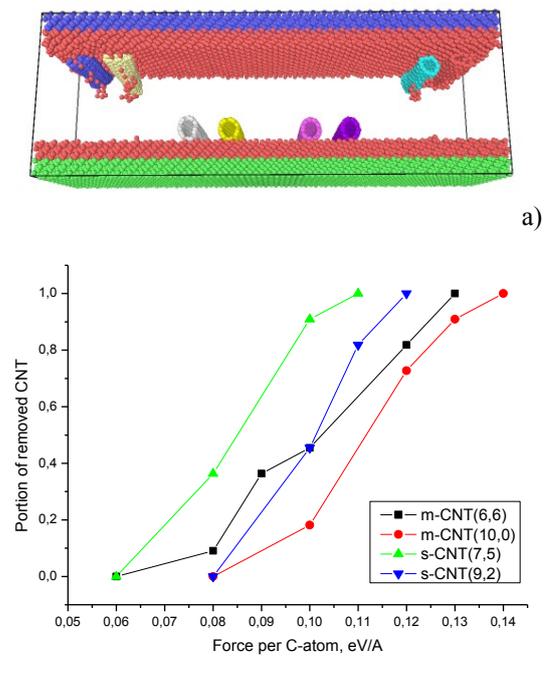

Figure 4. Examples of output data for MD simulations of the nanosubstrates functionalized by carbon nanotubes: (a) 3D visualization by AtomEye: nanotubes with various adhesion properties, (b) post-processing by R environment: the portion of removed nanotubes vs. the force applied to the nanotube

TABLE II. EXAMPLES OF SCALE LEVELS IN THE USE CASE OF MOLECULAR DYNAMICS SIMULATIONS OF NANOMATERIALS

| Level | Duration | Student Work | Topic |
|---|---|---|---|
| Nano (skill) | 10-30 min. | Input values | Basics of simulation (LAMMPS, NAMD, GROMACS, debyer), visualization (AtomEye, Ovito), simple plotting (R, ipython) |
| Micro (labwork) | 1-8 hours | Input script | Scenarios of MD simulations (LAMMPS, NAMD, GROMACS), statistical processing (R), complex plotting (R, ipython) |
| Mini (module) | 1-14 days | Set of scripts + parameters | Simple experiment (set of modules linked by scripts) |
| Macro (course) | 1-6 months | Tree of scripts + parameters | Complex experiment, even up to diploma work |

The additional way for variability of this approach can be illustrated by possibility to change the scale of the learning track. It is schematically depicted by connecting lines between modules in Fig. 1. The various scale levels are illustrated in Table 2 for the use case of molecular dynamics simulations of functionalized nanomaterials. All of these scale levels are actually constructed around various numbers of crates (SG components) and connecting links between them, which can be easily arranged and tuned in SG framework [16]. Our SG (http://scigate.imp.kiev.ua) in collaboration with High Performance Computing Center (http://hpcc.kpi.ua) was aimed on the close cooperation among the main actors in learning and researching world (teachers, students, scientists, supporting personnel, volunteers, etc.) with industry and academia to propose the new frameworks and interoperability requirements for the building blocks of a digital ecosystem for learning (including formal, non-formal, and informal learning) that develops and integrates the current and new tools and systems. This multi-level approach improves the effective study and research through flexible selection and combination of unified modules.

## V. CONCLUSIONS

Usage of SG ideology for development and integration of real-world and user targeted digital technologies for learning will foster the academia to use them and stimulate the market for innovations in educational technologies. Industrial actors can create "supermarkets" of modules and courses, created by academia or in cooperation with academia, and provide "menus" to obtain the customized learning "meals" (learning courses) created by teachers or/and users. It could boost the formal learning (schools, universities, etc.), the life-long and vocational fields of learning, and, especially, in-formal learning.

The novelties and advantages of the proposed paradigm (in comparison to the numerous available online learning solutions) consist of:

- homogenization and smoother integration of the very heterogeneous available learning content from various institutions (by the standard requirements for the new modules and the wrappers of legacy modules),
- close approach to the real-life learning situations,
- dynamic content, i.e. some modules in the shape of Virtual Labs around the simulations or remote experimentations (if the latter will be found among the potential partners),
- flexible combinations of various modules on demands of end users,
- modules with various versions differed by complexity (beginner, advanced, professional, etc.), scale (duration) like "nano" (10-30 min), "micro" (1-8 hours), "mini" (1-14 days), "macro" (1-6 months), and range,
- "billing" system around modules to provide the feedback for teachers about the quality, value, and necessity of some parts of the course or combinations of some modules,
- open or commercial repositories of the "bricks": modules, Virtual Labs, remote experimentations,
- portal(s) for content, tools, and user management.

This flexible and multiscale approach can help to implement the paradigm of ubiquitous learning (which development is usually restricted in time and space), and it can be feasible and efficiently implemented in the modern education system.

ACKNOWLEDGMENT

The authors would like to thank Prof. P.Kacsuk, D.G.Iglesia and A.Pester for fruitful discussions of this approach and valuable critics.